\documentclass[aps,pre,twocolumn,groupedaddress,showpacs,floatfix]{revtex4}
\usepackage{amsmath}
\usepackage{amssymb}
\usepackage{graphicx}

\begin{document}

\title{The L-Percolations of Complex Networks} 

\author{Luciano da Fontoura Costa}
\affiliation{Institute of Physics of S\~ao Carlos. 
University of S\~ ao Paulo, S\~{a}o Carlos,
SP, PO Box 369, 13560-970, 
phone +55 162 3373 9858,FAX +55 162 3373
9879, Brazil, luciano@if.sc.usp.br}

\date{\today}

\begin{abstract}   

Given a complex network, its \emph{L-}paths correspond to sequences of
$L+1$ distinct nodes connected through $L$ distinct edges.  The
\emph{L-}conditional expansion of a complex network can be obtained by
connecting all its pairs of nodes which are linked through at least
one \emph{L-}path, and the respective conditional \emph{L-}expansion
of the original network is defined as the intersection between the
original network and its \emph{L-}expansion.  Such expansions are
verified to act as filters enhancing the network connectivity,
consequently contributing to the identification of communities in
small-world models.  It is shown in this paper for $L=2$ and 3, in
both analytical and experimental fashion, that an evolving complex
network with fixed number of nodes undergoes successive phase
transitions -- the so-called \emph{L-}percolations, giving rise to
Eulerian giant clusters.  It is also shown that the critical values of
such percolations are a function of the network size, and that the
networks percolates for $L=3$ before $L=2$.

\end{abstract}


\maketitle

\section{Introduction} \label{sec:intr}

One of the most remarkable properties of complex networks is their
tendency to undergo a topological phase transition (i.e. percolation)
as the number of connections is progressively increased
\cite{Albert_Barab:2002,Newman:2003,Dorog_Mendes:2002}.  Several
systems, such as the world wide web and epidemics dissemination, only
become particularly interesting near or after percolation. Although
some works have investigated phase transitions in dynamical systems
underlain by complex networks
(e.g.\cite{Mulken_etal:2001,Sanchez:2001}), here we focus our
attention on \emph{topological} critical phenomena in such networks.
In addition to the classical studies by Erd\H{o}s and R\'enyi, more
recent works addressing percolation in networks include the analysis
of the stability of shortest paths in complex networks
\cite{Dong:2002}, the investigation of percolation in autocatalytic
networks \cite{Jain_Krishna:2000}, and the study of the fractal
characterization of complex networks \cite{Equiluz_etal:2003}.

\begin{figure}
 \begin{center} \includegraphics[scale=.5,angle=-90]{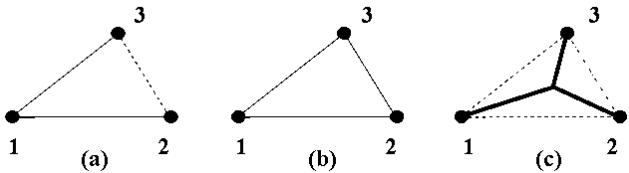} \\
   \caption{A pair of subsequent edges defining a \emph{virtual link}
   between nodes 2 and 3 (a), the basic cycle of size 3 underlying
   \emph{2-}percolations (b) and its relationship with the concept of
   \emph{hyperedge}(c).~\label{fig:threetri}} \end{center}
\end{figure}

The concept of \emph{L-}percolations is based on the
\emph{L-}expansions and \emph{L-}conditional expansions of a complex
network, which are introduced in the current paper. Consider the
two-subsequent edges in Figure~\ref{fig:threetri}(a).  The fact that
node $2$ is indirectly connected to node $3$ through a self-avoiding
\emph{2-}path passing through node $1$ defines a \emph{virtual link}
between nodes $2$ and $3$ \cite{Costa_what:2003}, represented by the
dotted line in that figure.  In case such a virtual edge does belong
to the network, it defines a cycle of length $3$ between the three
involved nodes, as illustrated in Figure~\ref{fig:threetri}(b), which
contains three distinct undirected \emph{2-}paths.  The
`self-avoiding' requirement for the paths is imposed in order to avoid
passing twice through the same edge, which would tend to produce an
infinite number of paths between any pair of connected edges.  For
instance, in case this restriction is not considered, there would be
an infinite number of even-length paths (i.e. $L = 3, 5, 7, \ldots$)
from node 1 to 2 in Figure~\ref{fig:threetri}(a), while just the
direct \emph{1-}path between those nodes remain after imposing the
self-avoiding requirement.  Such a restriction therefore allows a
clearer characterization of the distribution of path lengths and
connectivity in a complex network.

The specific demands governing the network growth may imply that
cycles such as that in Figure~\ref{fig:threetri}(b) occur sooner or
later along the network evolution.  For instance, intense indirect
(i.e. through node $1$) information exchange between nodes $2$ and $3$
is likely to foster the appearance of the direct link between those
two nodes \cite{Costa_what:2003}.  In other words, the formation of
such cycles can be understood as a \emph{reinforcement of the
connectivity between the involved nodes}, possibly implied by
intensive information interchange and/or node affinity.  Therefore,
the density of \emph{3-}cycles (as well as cycles similarly defined
for other values of $L$) is likely to provide interesting insights
about the growth dynamics and connectivity properties of complex
networks.  This is one of the main motivations for the investigations
reported in the present article, with implications to the
identification of communities in the analyzed networks, as described
below.  Another important aspect intrinsic to the definition of
\emph{L-}paths is that such a property is directly related to the
\emph{transitivity} of connections along the network.  In other words,
if node $i_1$ is connected to node $i_2$, which is connected to node
$i_3$, and so on, until $i_L$, the eventual presence of the therefore
established virtual link extending from $i_1$ to $i_L$ can be
understood as an indication of transitivity in the network
connectivity.

It is worth observing that a \emph{3-}cycle can also be related to a
\emph{hyperedge} between the three involved nodes, as illustrated in
Figure~\ref{fig:threetri}(c).  Such a kind of edge is characteristic
of \emph{hypergraphs} \cite{Bollobas:2002, West:2001, Chartrand:2000},
whose nodes are connected through hyperedges.  In other words, a
hyperedge is a relationship defined at the same time between more than
two nodes. The essential difference between the \emph{3-}cycle shown
in Figure~\ref{fig:threetri}(b) and the hyperedge in (c) is the fact
that the \emph{3-}cycle can be dismantled by removing any of the
involved edges while the hyperedge implies simultaneous deletion of
all connections (i.e. a hyperedge is a single edge and can only be
removed as a whole).  In other words, a \emph{3-}hyperedge
intrinsically implies a \emph{3-}cycle, but not the other way round.
Still, situations where the \emph{(L+1)-}cycles are closely
interdependent can be understood in terms of hyperedges.

\begin{figure}
 \begin{center} 
   \includegraphics[scale=.5,angle=0]{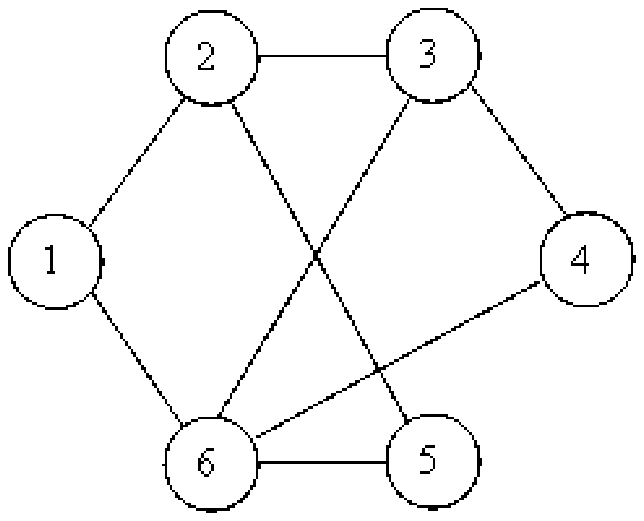} \\
   (a)  \\
   \includegraphics[scale=.45,angle=0]{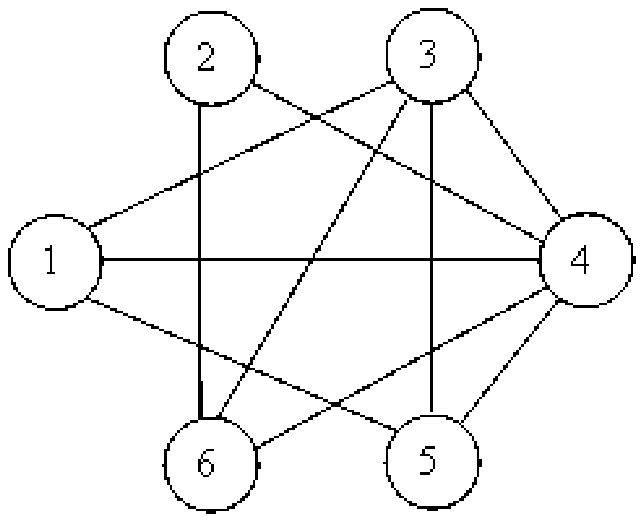} 
   \includegraphics[scale=.45,angle=0]{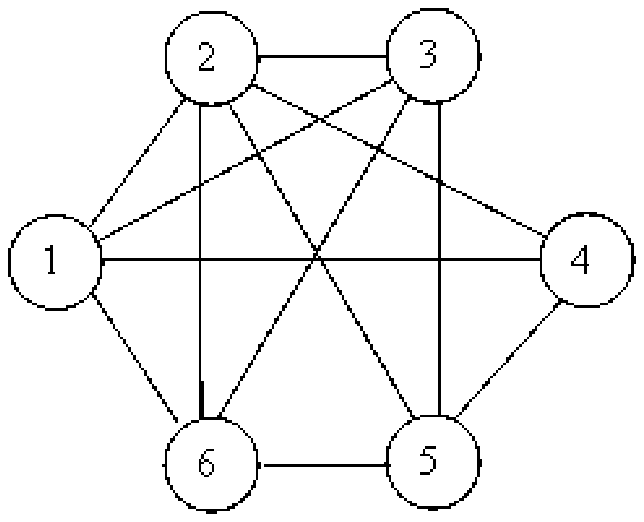} \\
   (b) \hspace{3cm}  (c) \\
   \includegraphics[scale=.45,angle=0]{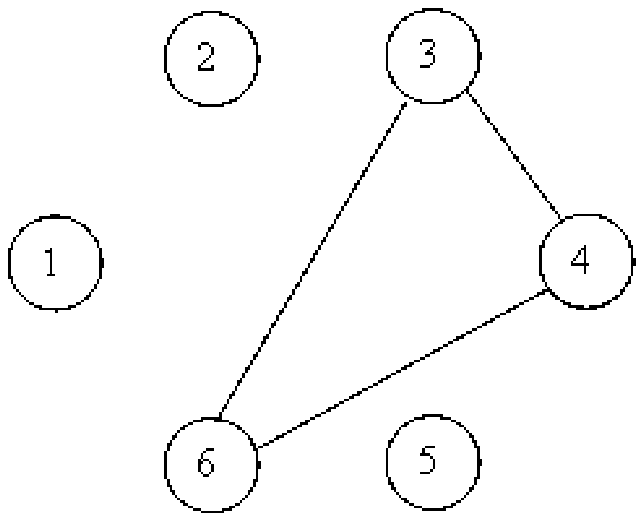} 
   \includegraphics[scale=.45,angle=0]{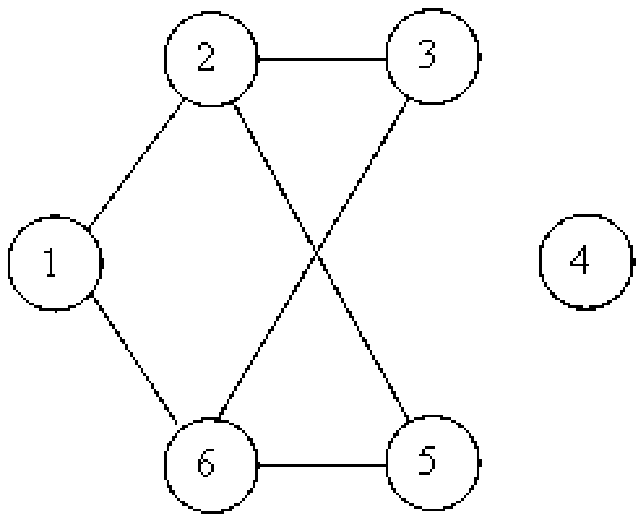} \\
   (d) \hspace{3cm}  (e) 

   \caption{A simple graph (a) and its respective 
   \emph{2-} (b) and \emph{3-}expansions (c).  The respective
   \emph{2-}  and \emph{3-}conditional expansions are shown in (d) and 
   (e).  The completeness ratios are  $f_2=0.43$ and 
   $f_3=0.86$.~\label{fig:ex_contr}} 
 \end{center}
\end{figure}

Given a complex network $\Gamma$ with $N$ nodes, its
\emph{L-}expansion is henceforth defined as the new network, also with
$N$ nodes, which is obtained by incorporating an edge between two
nodes $i$ and $j$ whenever there exists a virtual link of length $L$
(i.e. a self-avoiding path of length $L$) between those two nodes.
Observe that such a definition holds for both directed and undirected
graphs.  Such expansions can be implemented by considering or not
multiple links between two nodes, leading to different results.  The
\emph{L-}expansion of a network can be intersected with the original
network $\Gamma$ in order to obtain the \emph{L-}conditional expansion
of that network.  Such a network, which also involves $N$ nodes,
contains all cycles of length \emph{L+1} in the original network and no
node with null or unit degree.  Actually, all connections in such a
network belong to cycles of length $L+1$.  In case $L=2$, the
respective \emph{L-}conditional expansion is composed exclusively of
\emph{3-}cycles, and can therefore be related to a respective
\emph{3-}hypergraph contained in the original network.
Figure~\ref{fig:ex_contr} illustrates a simple undirected graph (a)
and its respective \emph{2-} (b) and \emph{3-}expansions (c). The
\emph{2-} and \emph{3-}conditional expansions of the graph in
Figure~\ref{fig:ex_contr}(a) are shown in (d) and (e).

\begin{figure}
 \begin{center} 

   \includegraphics[scale=.4,angle=-90]{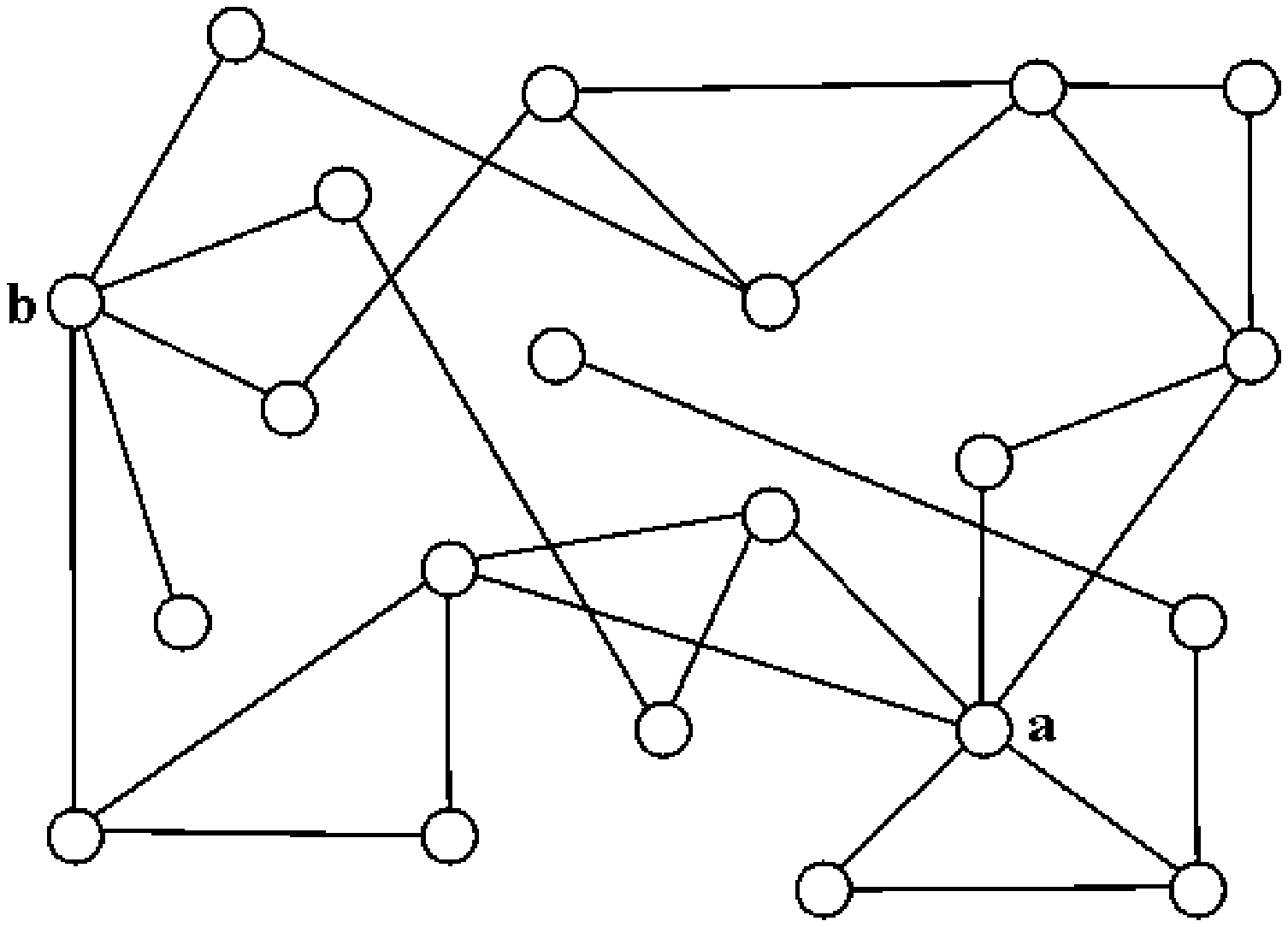} \\ (a) \\
   \includegraphics[scale=.4,angle=-90]{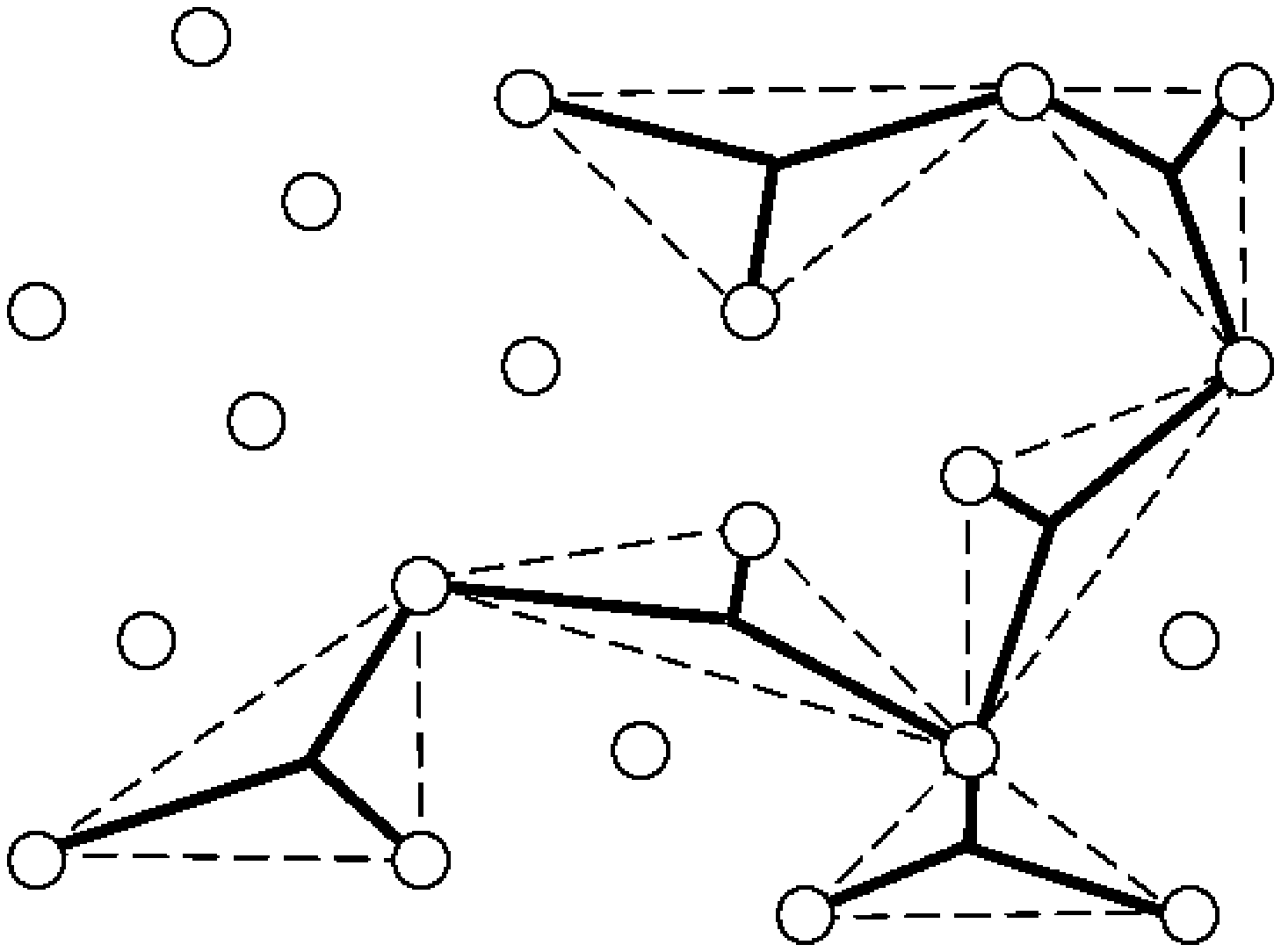} \\ (b)
   \caption{A simple network (a) and its \emph{2-}conditional
   expansion represented in terms of
   cycles/hyperedges.~\label{fig:kernel}}

\end{center}
\end{figure}

Figure~\ref{fig:kernel} shows a simple network (a) and its respective
\emph{2-}conditional expansion, which involves only \emph{3-}cycles.
Observe that the conditional percolation removes all the edges not
involved in \emph{3-}cycles, so that the obtained clusters represent a
strong backbone of the original network.  Observe that any single edge
can be removed from the conditionally-expanded network without
producing a new cluster, which is true for any $L \geq 2$.  At the
same time, the conditional expansion tends to enhance the clustering
coefficient \cite{Albert_Barab:2002} of the obtained network.  For the
specific case $L=2$, the conditional expansion also tends to enhance
the \emph{regularity}
\footnote{A network is regular iff all its nodes have the same
degree.} of the network, in the sense that all nodes in the resulting
network have degree equal to an integer multiple of $2$.

Conditional expansions can be thought of as a kind of network filter
which removes edges in order to preserve those groups of nodes more
densely connected through \emph{L-}paths, therefore characterizing a
tendency of the conditional expansion to preserve the subgraphs
contained in small-world models \cite{Watts_Strog:1998,
Albert_Barab:2002}.  This fact makes the conditional expansions an
interesting mechanism for identifying communities in such a kind of
complex networks, a possibility which is preliminarly investigated in
the current work (see Section~\ref{sec:comm}).  Observe that the hubs,
i.e. nodes with high degree are not necessarily preserved by
conditional expansions.  For instance, although the hub marked as
\emph{a} in Figure~\ref{fig:kernel} was retained, the one marked as
\emph{b} was eliminated by the \emph{3-}conditional expansion.

The progressive addition of new edges into a random network inevitably
leads to the appearance of a \emph{giant cluster}, which dominates the
network henceforth.  Such a phenomenon, corresponding to a topological
phase transition (i.e. \emph{percolation}) \cite{Stauffer_Aha:94} of
the network, has motivated much interest and bridged the gap between
graph theory and statistical mechanics.  As shown in this article both
in analytical and experimental fashion, the uniform evolution of a
random network (and also of a preferential-attachment model) naturally
leads not only to the traditional percolation, but also to successive
\emph{L-}percolations, which are characterized by the fact that a
giant cluster appears in the respective \emph{L-}conditional
expansions.  For instance, as the number of edges is progressively
increased, one reaches a point where a giant cluster is established in
the original network where each of its nodes is connected to at least
another node not only through a direct connection, but also through a
self-avoiding path of length $L$.  This is the \emph{main property}
characterizing the giant clusters for generic \emph{L-}percolations.
For instance, in case $L=2$, the giant cluster is characterized by the
fact that each of its edges is part of a \emph{3-}cycle.  In other
words, every edge of the giant cluster can be associated to a
\emph{3-}hyperedge, thus emphasizing the fact that the connections of
such a \emph{2-}giant cluster are stronger than the traditional case
(i.e. $L=1$).  The hypergraph in Figure~\ref{fig:kernel} provides a
simplified illustration of the connections characterizing the giant
cluster in a \emph{2-}percolated complex network.  It should be
noticed that the giant cluster obtained in an \emph{L-}conditional
expansion has the unique characteristic that it can not be modified by
a subsequent conditional expansion for the same value $L$, a property
called \emph{idempotency}.

The concept of \emph{L-}percolations allows a series of insights about
the analyzed networks, including:

\emph{Theoretical features:} As one of the main interests in complex
networks is related to the occurrence of critical topological
transitions, the identification of further percolations of a growing
network represents an important fact by itself, indicating that the
abrupt changes of the network properties, namely connectivity, is not
restricted to 1-connectivity, but extends over several values of $L$.
Therefore, the dynamics of network connectivity formation is verified
to be richer than usually considered in complex network theory.

\emph{Identification of the strongest connections:} The occurrence of
an \emph{L-}percolation indicates a reinforcement between most of the
connections in a network as far as \emph{L-}paths are concerned.  In
other words, the fact that a giant cluster is obtained for a specific
value of $L$ indicates that most of the network nodes are connected
through closed \emph{L-}paths (i.e. most belong to cycles), therefore
exhibiting enhanced connectivity for that value, in the sense that
several links may be removed before the percolated cluster collapses.

\emph{Network characterization:} Several measurements related to the
concept of \emph{L-}connectivity can be defined and used to
characterize the connectivity properties of complex networks (such as
transitivity), as illustrated in the last Section of this article.
Taken in combination with traditional features such as the average
node degree \cite{Albert_Barab:2002, Newman:2003, Dorog_Mendes:2002},
such measurements allow a more complete characterization of the
connectivity features of the network under analysis.

\emph{Community finding:} The clusters obtained by the
\emph{L-}conditional expansions are characterized by enhanced average
clustering coefficient while presenting a tendency to remove links
between loosely connected groups of nodes, suggesting the division of
the original network into meaningful communities
(e.g. \cite{Newman_Girvan:2004, Capocci:2004}).  In other words, the
\emph{L-}conditional expansions of a network may help the
identification of the involved communities, as only the more intensely
connected nodes (i.e. those characterized by \emph{L-}paths) will
remain after the conditional expansion. This possibility is
preliminarly illustrated in Section~\ref{sec:comm} of this article.

\emph{Eulerian networks:} In case the conditional expansions are
performed allowing for multiple links between two nodes, it is shown
in this paper that the giant clusters underlying the
\emph{L-}percolations, and therefore contained in the original
network, are necessarily Eulerian \footnote{A connected graph is
\emph{Eulerian} if it contains a closed trail including all edges,
i.e. an \emph{Eulerian trail}.  A \emph{trail} is a sequence of
distinct edges connecting not necessarily distinct nodes.}.  In other
words, all nodes of the giant cluster defined by an
\emph{L-}percolation can be visited without passing twice over the
same edge. Such a property has interesting implications for several
types of complex networks, including communication networks and
protein sequence networks (e.g. \cite{Hao_Eulerian:2001}), as well as
for random walk investigations \cite{Povolotsky:1998}.

\section{Definitions}

Let the nodes of the network of interest $\Gamma$ be represented as $k
= 1, \dots, N$, and the edges as ordered pairs $(i,j)$.  The total
number of edges in the network is expressed as $n(\Gamma)$.  For
generality's sake, the network $\Gamma$ can be fully represented by
its weight matrix $W$, where no self-connections are allowed. The
weight matrix is obtained by assigning the weight of each edge $(i,j)$
to the respective weight matrix element $w(j,i)$.  Such a
representation of the network in terms of a weight matrix is more
general than the traditional adjacency matrix, as it allows the
representation of the weights associated to the network edges.  To any
extent, the adjacency matrix $A$ of a graph can be obtained by making
its elements $a(j,i)=1$ whenever there is an edge (of any non-zero
weight) between nodes $i$ and $j$.  The average node degree of
$\Gamma$ is henceforth represented as $z$.  Let $\delta_T(x)$ be the
operator acting elementwise over the matrix $x$ in such a way that the
value one is assigned whenever the respective element of $x$ has
absolute value larger or equal than the specified threshold $T$; for
instance, $\delta_2(\vec{x}=(3 ,-2, 0, -4, 1))=(1,1,0,1,0)$.  Thus,
the adjacency matrix can be obtained from the weight matrix as
$A=\delta_T(W)$.

The random and preferential-attachment models were obtained as
described in the following.  For the random network case, the growing
parameter corresponds to the Poisson rate (the mean density of
connections) $p$, and in the preferential attachment case, to the
number $c$ of existing connections in the current stage of growth,
which are normalized in terms of the respective average node degree as
$z_R=p (N-1)$ and $z_{SF}=2c/N$ respectively, for the sake of direct
comparison between the two network types.  The preferential-attachment
network growth was performed by keeping a list where each network node
with current degree $k$ appears $k$ times, so that subsequent edges
can be defined by selecting pairs of distinct elements in such a list
according to the uniform distribution \cite{Stauffer:2003}.  
 
One can check for an \emph{L-}path between two nodes $i$ and $j$ by
checking the sequences of successors for each transition along the
graph, until either a path is found or all the paths have been tested
(the so-called \emph{depth-first} scanning of the graph from node
$i$).  Only successor nodes not already included in the constructing
path are considered during this procedure.  Observe that, in the
particular case of $L=2$, the respective paths can be immediately
obtained from the squared adjacency matrix ignoring its main diagonal.

The \emph{L-}expansions and \emph{L-}conditional expansions are
henceforth represented as $\Upsilon_L$ and $\Xi_L$, respectively.
Interesting information about the intrinsic topology of the original
network $\Gamma$ can be obtained from their measurements of the
respective expansions.  Some simple possibilities are the number of
edges in each of these graphs, henceforth expressed as $n(\Upsilon_L)$
and $n(\Xi_L)$.  Thus, $f_L=n(\Xi_L)/n(\Gamma)$, hence the
\emph{completeness ratio} of the network $\Gamma$ for $L$, corresponds
to the ratio between the number of edges belonging to $(L+1)-$cycles
and the total number of edges in the original network $\Gamma$.  For
instance, in case all edges in $\Gamma$ are part of a \emph{3-}cycle, we
have $f_2=1$, while smaller values are obtained for less transitive
networks.   

\begin{figure}
 \begin{center} 

   \includegraphics[scale=.55,angle=-90]{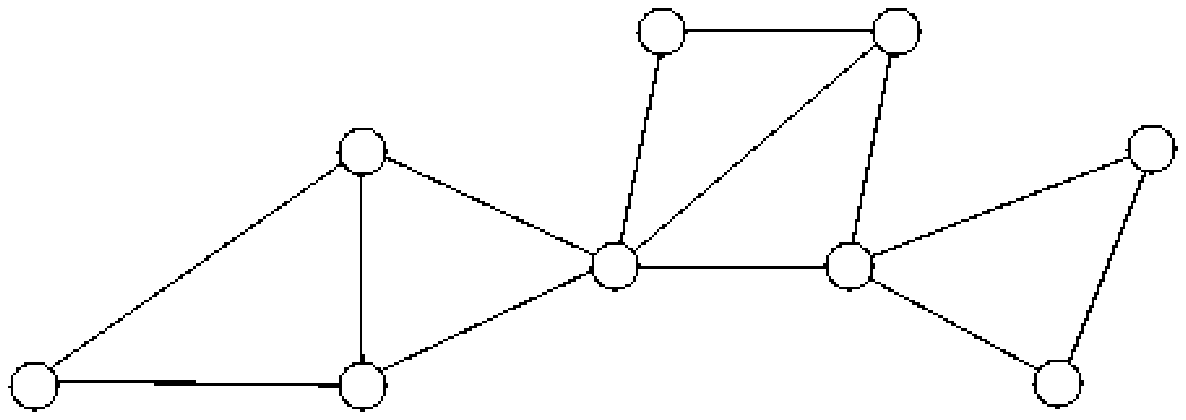} \\ (a) \\
   \includegraphics[scale=.55,angle=-90]{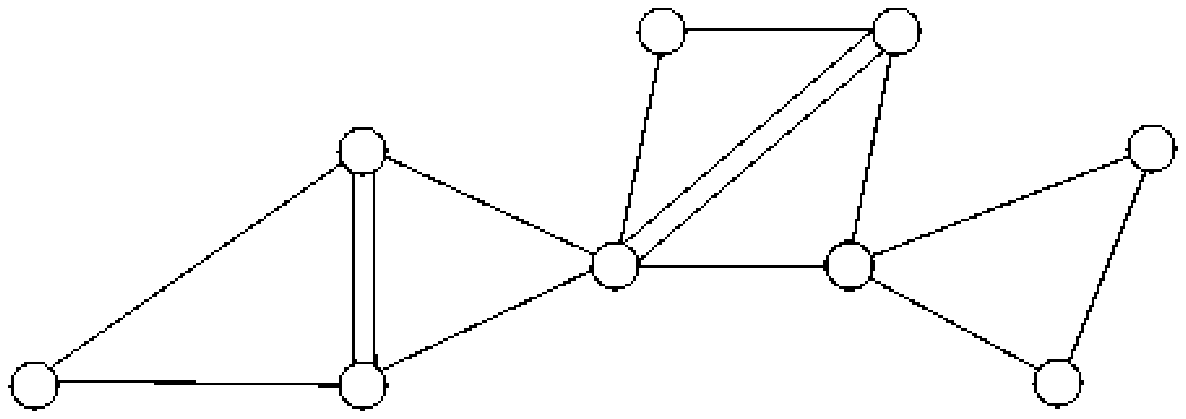} \\ (b)
   \caption{A simple network obtained after \emph{2-}conditional
   expansion (a) and the duplication of its shared edges in order to
   warrant the Eulerian property.~\label{fig:euler}}

\end{center}
\end{figure}

An interesting property of the giant clusters defined by the
\emph{L-}percolations when multiple links are allowed is the fact that
they are necessarily Eulerian.  More specifically, such multiple edges
are implemented as follows: given the \emph{2-}conditional expansion
of the original network, duplicate each edge which is shared by two
cycles, as illustrated in Figure~\ref{fig:euler}.  This interesting
property can be easily proved by considering that all nodes in such a
network will have even degree, which is a necessary and sufficient
condition for the cluster to be Eulerian \cite{Chartrand:2000}.

\section{Conditional Expansions and Community Finding} \label{sec:comm}

\begin{figure}
 \begin{center} 

   \includegraphics[scale=.7,angle=-90]{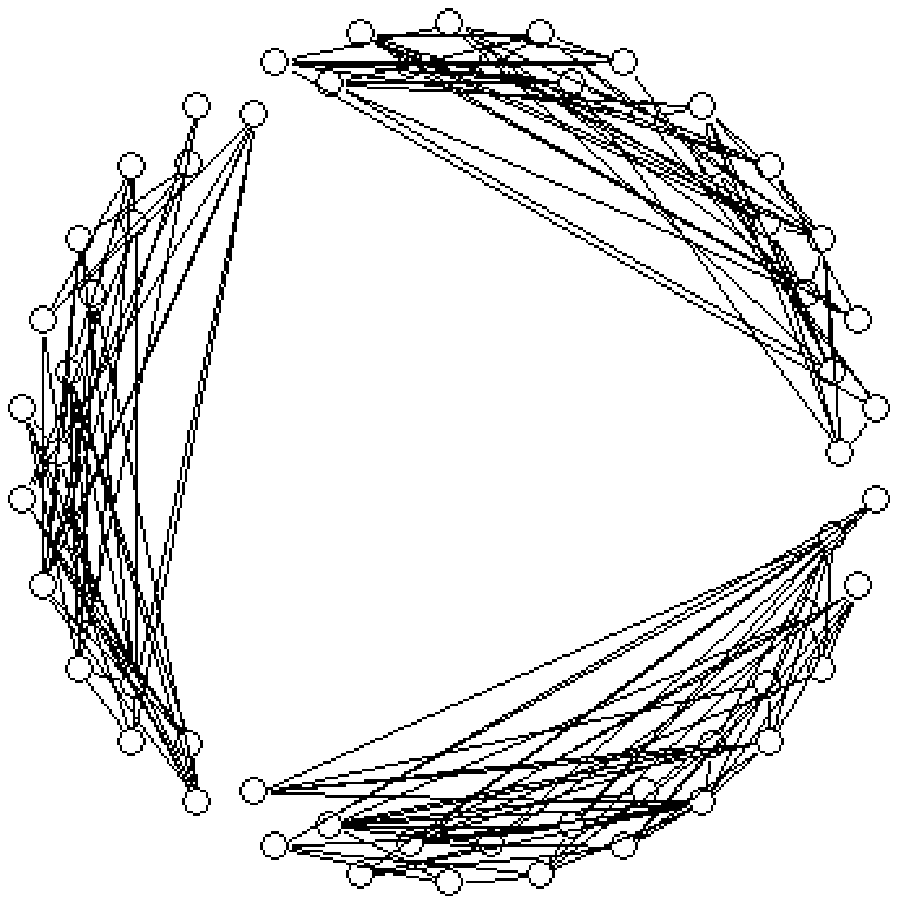} \\ (a) \\
   \includegraphics[scale=.7,angle=-90]{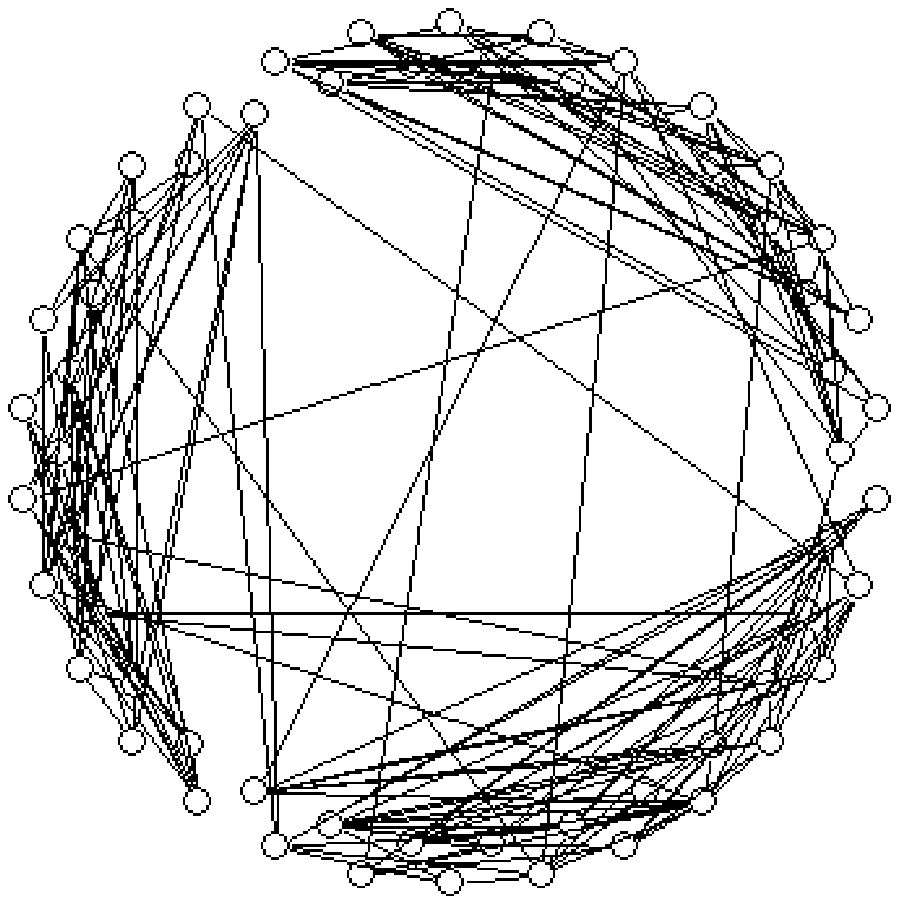} \\ (b) \\
   \includegraphics[scale=.7,angle=-90]{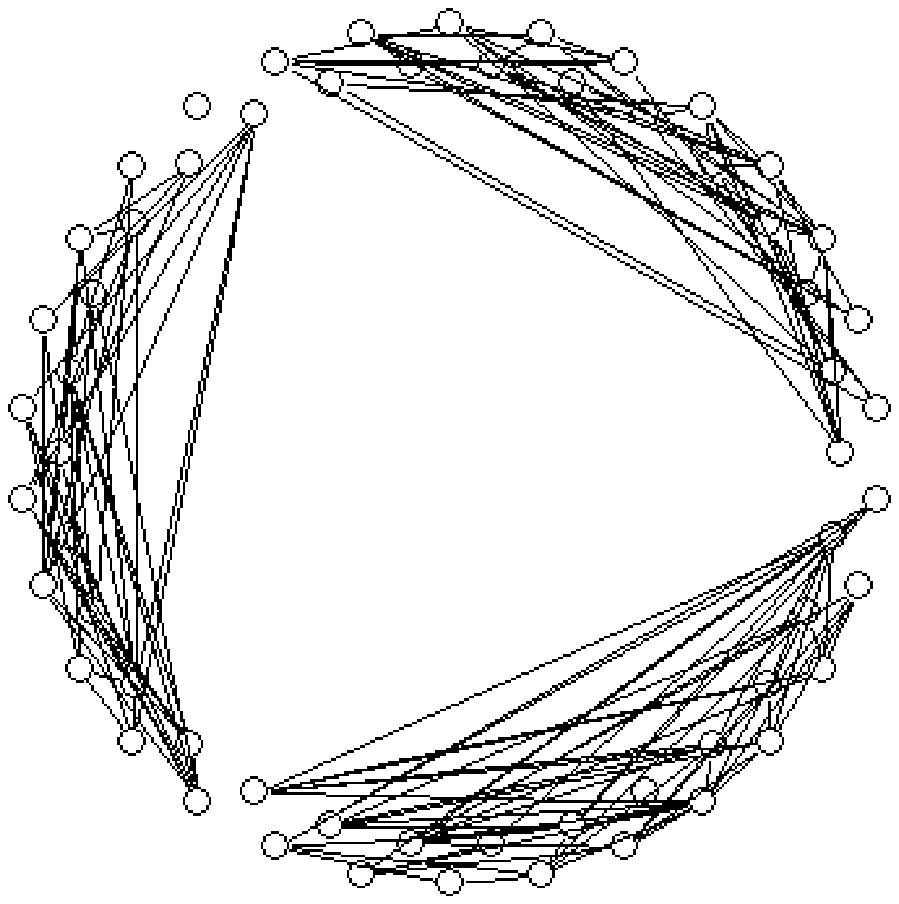} \\ (c)
   \caption{Illustration of how the conditional expansion of a network
   can act as a filter enhancing the communities in small-world types
   of networks: (a) original network containing three communities; (b)
   network obtained by addition of random connections between
   communities; and (c) partial identification of the three original
   communities obtained by \emph{2-}conditional expansion of the
   network in (b).~\label{fig:comm}}

\end{center}
\end{figure}

\begin{figure}
 \begin{center} 
   \includegraphics[scale=.7,angle=-90]{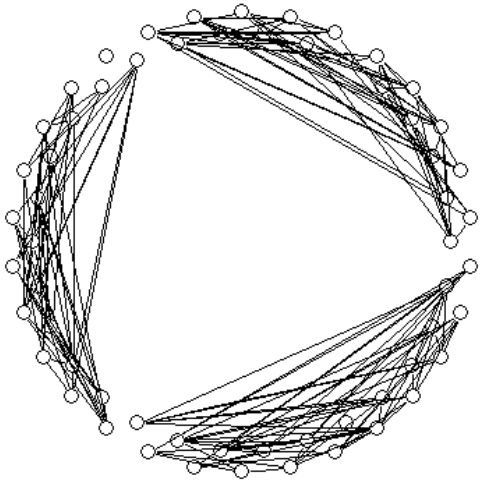} \\
  
   \caption{Complemented communities obtained by incorporating edges
   from the network in Figure~\ref{fig:comm}(b) into the three
   respective clusters in Figure~\ref{fig:comm}(c).~\label{fig:comm2}}
\end{center}
\end{figure}

This section reports a preliminarly investigatation of the potential
of \emph{L-}conditional expansions as a subsidy for community finding.
We already observed in Section~\ref{sec:intr} of this article that
such expansions tend to preserve the more strongly connected
subnetworks or communities.  Figure~\ref{fig:comm}(a) shows an initial
network containing 3 communities of 20 nodes, each corresponding to a
random network whose edges were added according to uniform probability
	with Poisson rate 0.3.  By adding random edges between these 3 groups,
small-world networks such as that shown in Figure~\ref{fig:comm}(b)
can be obtained.  The potential of the \emph{L-}conditional expansions
for isolating the original communities is illustrated in
Figure~\ref{fig:comm}(c), which shows the network obtained by
performing the \emph{2-}conditional expansion over the network in
Figure~\ref{fig:comm}(b).  Although missing some of the original
connections, the three clusters obtained by the conditional expansion
correspond to the original communities.  The connections inside each
obtained community can be partially complemented by incorporating
those edges in Figure~\ref{fig:comm}(b) which connect nodes inside
each of the identified clusters.  This type of link is henceforth
called an \emph{intracommunity} edge, while links between nodes in
different communities are called \emph{intercommunity} edges. The
effect of this procedure is illustrated in Figure~\ref{fig:comm2} with
respect to the three communities identified in
Figure~\ref{fig:comm}(c).  Although a total of 19 edges were
recovered, the disconnected node at the upper left corner of the
network remained isolated from its original community.

We have verified through experimental simulations that, for a
relatively small number of added edges (Poisson rate about 0.015 for
networks such as that in Figure~\ref{fig:comm}), the original
communities can be reasonably estimated by applying the
\emph{2-}conditional expansion.  Situations where some intercommunity edges
are left by such expansions can be addressed by using the Newman and
Girvan's method after the respective \emph{2-}conditional expansion.
However, situations involving a substantial number of random
intercommunity edges, as when the intercommunity and intracommunity
edges are established with similar probabilities, are likely to
produce incorrect communities.

\section{\emph{L-}Percolations}

The \emph{L-}percolations of random networks have been investigated
from both the analytical and experimental points of view, as described
in the following.  Preferential-attachment networks were also
considered, but only experimentally.

\subsection{Analytic Mean-Field Calculations}

The following mean-field approximation assumes that the edge
assignment takes place independently of the node degree or other
network features, being more accurate for not extremely dense networks
(see below).  Let $\Gamma$ be a random network with $N$ nodes and
Poisson rate $p$ (defining the overall edge density), implying the
maximum number of edges to be $n_T=N(N-1)/2$. It immediately follows
that the average number of expected edges is $n=pn_T$ and that the
average node degree therefore is $z=2n/N=p(N-1)$. We adopt a mean
field approach considering each node $i$.  As the expected number of
nodes connected to $i$ is $z$, we have that the maximum number of
\emph{2-}paths involving the edges $(k,i); (i,p)$ established between
any two nodes $k$ and $p$ connected to node $i$ is $z(z-1)/2$, so that
the expected number of direct connections between them is $pz(z-1)/2$
and the expected total number of \emph{2-}paths is $n_2=Npz(z-1)/2$,
implying the respective \emph{2-}conditional expansion of the original
graph to have Poisson rate $p_2$ as given by Equation~\ref{eq:p2}.
Now, since the \emph{2-}conditional expansion of $\Gamma$ is also a
random network (recall that the edge assignment takes place
independently of node degree or other current network property), we
obtain the critical point $p^c_2$ where the respective percolation
takes place by considering that the node degree of the
\emph{2-}conditional expansion, namely $z_2^c(2-ce)$, reaches unit
value at that critical transition (see Equation~\ref{eq:pc2}).
Interestingly, this critical point is verified to grow as a power of
$N$.  The respective average degree of the original network at this
critical point can be estimated as in Equation~\ref{eq:zc2}.

Next we address the case $L=3$ by considering each edge $e$ connecting
two nodes $i$ and $j$ of average degree $z$.  The total of direct
connections between the distinct nodes $k$ and $p$ connected
respectively to $i$ and $j$, defining \emph{3-}paths $(k,i); (i,j);
(j,p)$ between nodes $k$ and $p$, therefore is given as $z^2$, and the
expected number of such directed links per edge therefore is $(pz)^2$.
The total number of direct links between the pairs of nodes connected
through 3-paths passing through edge $e$ therefore is $n_3 = n_T(pz)^2
= Np^4(N-1)^3/2$, implying respective Poisson rate as in
Equation~\ref{eq:p3}.  Observe that, in case the connections become
too dense, the number of closed paths such as $(k,i); (i,j); (j,k)$,
instead of the assumed open paths $(k,i); (i,j); (j,p)$, may become
too frequent and undermine the above estimations for $L=3$.  As the
\emph{3-}conditional expansion of the original random network is also
assumed to be a random network, the associated critical rate value can
now be calculated as in Equation~\ref{eq:pc3}.  The corresponding
average degree of the original network at this critical point is given
by Equation~\ref{eq:zc3}.  Interestingly, it follows that $p^c_3 <
p^c_2$ for any $N>1$, i.e. the 3-conditional expansion is expected to
percolate sooner than the \emph{2-}conditional expansion.  We also
have that $f_2=p^2(N-1)$, $f_3=p^3(N-1)^2$, and $f_3/f_2=p_3/p_2=z$.

\begin{eqnarray}
  p_2 = \frac{n_2}{n_T} = p^3(N-1)  \label{eq:p2} \\
  z^c_2(2-ce) = 2 \frac{n_2}{N} = 1 
     \Longrightarrow p^c_2 = (N-1)^{-2/3}  \label{eq:pc2} \\
  z^c_2(orig) = p^c_2 (N-1) = (N-1)^{1/3} \label{eq:zc2} \\
  p_3 = \frac{n_3}{n_T} = p^4(N-1)^2  \label{eq:p3} \\
  z^c_3(3-ce) = 2 \frac{n_3}{N} = 1 \Longrightarrow 
     p^c_3 = (N-1)^{-3/4}  \label{eq:pc3}  \\
  z^c_3(orig) = p^c_3 (N-1) = (N-1)^{1/4} \label{eq:zc3}
\end{eqnarray}

\subsection{Simulation Results}

\begin{figure*}
 \begin{center} \includegraphics[scale=.72,angle=-90]{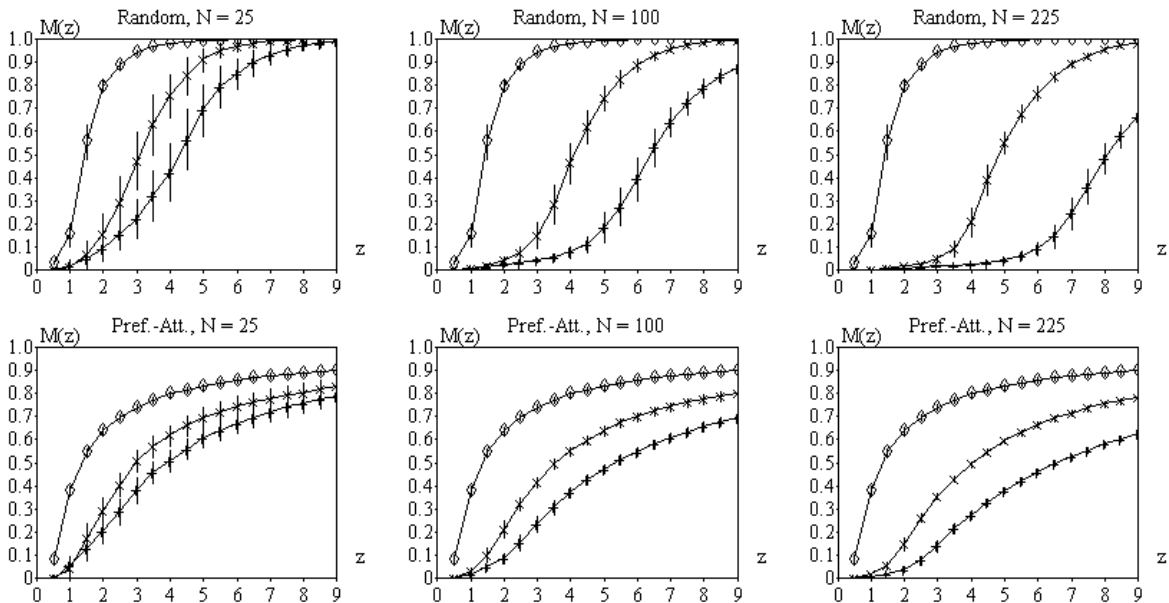} \\

   \caption{The giant cluster sizes for the random and preferential
   attachment networks and respective conditional expansions in terms
   of $z$ for $N=25, 100$ and $225$ and $L=1$ ($\diamond$), $L=2$
   ($+$) and $L=3$ ($\times$).~\label{fig:res}}

\end{center}
\end{figure*}

\begin{figure}
 \begin{center} \includegraphics[scale=.6,angle=-90]{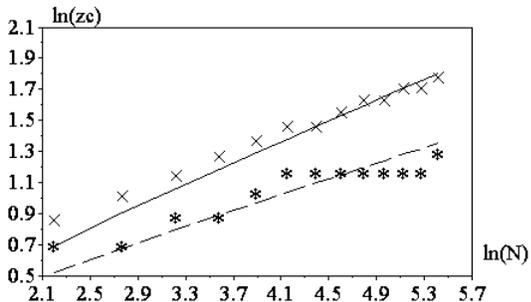} \\

   \caption{Dilog presentation of analytical predictions (solid line:
   $L = 4$; dashed line: $L = 3$) and experimental values ($\times$:
   $L = 2$; $*$: $L = 3$) of critical average node degrees in
   terms of $N$.~\label{fig:expana}}

\end{center}
\end{figure}

The experimental analysis was performed by monitoring the normalized
maximum cluster size $M(z)$ for each instance of the growing networks
for a total of 300 realizations of each configuration.
Figure~\ref{fig:res} shows the maximum cluster sizes obtained for
random and preferential attachment networks for $N=b^2$, where $b = 3,
4, \ldots, 15$.  The preferential attachment cases where generally
characterized by smoother transitions than their random counterparts.
As expected, the \emph{2-} and \emph{3-}conditional expansions
percolated later than the original network ($L=1$), with the
\emph{3-}conditional expansions of the original network percolating
sooner and more abruptly than the respective \emph{2-}conditional
expansions.  Larger dispersions were observed for the maximum cluster
sizes of the random networks, indicating that their specific
realizations are less uniform.  As theoretically predicted, the
critical average node degree $z_c$ resulted a function of the network
size $N$ in both the random and preferential-attachment models, but
such a dependency was markedly less intense for the
preferential-attachment model.  The dilog diagram in
Figure~\ref{fig:expana} shows the theoretical predictions (solid and
dashed lines) and the values corresponding to 80\% of the critical
average node degrees.  More specifically, the vertical axis represents
the logarithm of the critical average node degree corresponding to
80\% of the value of $z$ for which the maximum dispersion of $M(z)$
(the vertical bars in Figure~\ref{fig:res}) is observed in the
respective simulation.  Observe that a total of 13 simulation sets
(executed for 200 realizations considering each specific values of
$N$, as identified in the x-axis of Figure~\ref{fig:expana}), and not
only the three cases illustrated in the first row of
Figure~\ref{fig:res}, were considered in order to obtain the results
shown in Figure~\ref{fig:expana}. A good agreement between analytic
and experimental results is verified for both $L=2$ and 3.

\begin{figure*}
 \begin{center} \includegraphics[scale=.73,angle=-90]{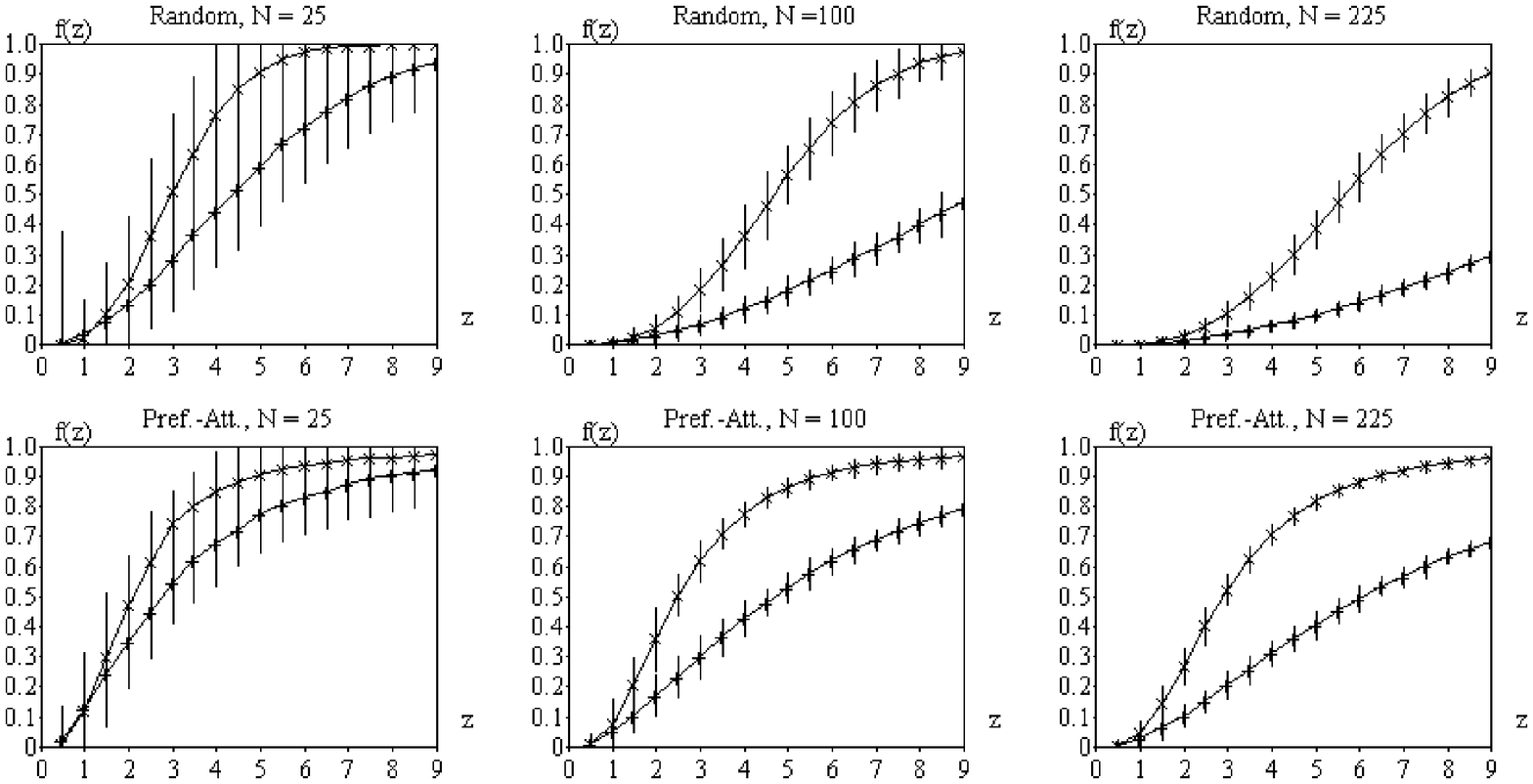} \\

   \caption{The completeness ratios for the random and
   preferential-attachment models in terms of $z$ for $N=25, 100$ and
   $225$ and $L=2$ ($+$) and $L=3$ ($\times$).~\label{fig:crat}}

\end{center}
\end{figure*}

The completeness ratios for the random and preferential-attachment
have also been estimated in our simulations, and the results for
$N=25, 100$ and $225$ are shown in Figure~\ref{fig:crat}.  It is clear
from this figure that, as expected, higher completeness ratios were
always obtained for the \emph{3-}conditional expansions than for the
\emph{2-}conditional expansions, substantiating the fact that
\emph{3-}conditional expansions tend to produce networks with more edges
than those obtained for \emph{2-}conditional expansions.  At the same time,
the preferential-attachment models were characterized by completeness
ratios which grow faster with $z$ than for the random model,
indicating that this type of complex network tends to produce a higher
number of edges belonging to \emph{3-} and \emph{4-}cycles.  Such a result is
compatible with the fact that random networks tend to percolate faster
than the preferential-attachment model because the completeness ratio
only takes into account the number of existing edges in the original
network.  For instance, a network containing 10 isolated \emph{3-}cycles
will have $f_2=1$ while a network with a single cycle containing 30
nodes will imply $f_2=0$.  Therefore, while the size of the dominant
cluster tend to grow slower in preferential attachment networks, the
clusters in such cases have a higher number of edges belonging to \emph{3-}
or \emph{4-}cycles than those in random networks, as expressed in
Figure~\ref{fig:crat}.

\section{Application to Real Data}

In order to illustrate the potential of the concepts and measurements
suggested above, they have been applied to experimental data regarding
concept associations obtained in the psychophysical experiment
described in \cite{Costa_what:2003}, where a human subject is
requested to associate words.  After 1930 word associations were
obtained, a weighted directed graph $\beta$ was determined by
representing each of the 250 words as nodes and the number of specific
associations between two words as the weight of the edge between the
respective nodes.  The respective average node seemed to follow a
power law.  Here we consider the adjacency matrix
$A=\delta_{T=1}(B+B')$, where $B=\delta_{T=1}(W)$, $B'$ is the
transpose of $B$ and $W$ is the weight matrix obtained in the concept
association experiment.  In other words, the original weight matrix is
thresholded at $T=1$ and made symmetric in order to transform the
original digraph into a graph.  Matrix $B$ was characterized by
$n=738$ edges, $p=0.011$ and $z=5.9$.  The \emph{2-} and
\emph{3-}conditional expansions of $A$ were obtained as described
above, leading to $n_2=407$ and $n_3=606$ connections, $p_2=0.0067$ and
$p_3=0.0092$, with respective average node degrees of 4.5 and 5.7.
Thus, we have $f_2=0.55$ and $f_3=0.82$ indicating, as expected, a
higher density of \emph{4-} than \emph{3-}cycles in the original
network.  Such high densities also imply that about half of the edges
in the original network belong to \emph{3-}cycles, while over 80\% of
the edges belong to \emph{4-}cycles.  As a matter of fact, the fact
that these values exceed the respective critical values $z_c$ for
$N=250$ (see Figure~\ref{fig:expana}) strongly suggest that the word
association network has already undergone percolation as well as \emph{2-}
and \emph{3-}percolations.  Such results indicate that the word association
network is underlain by \emph{3-} and \emph{4-}cycles formed as a
consequence of transitive nature of word associations.  

An interesting question implied by the above results regards the
quantification of the degree of transitivity in a complex network in
terms of the completeness ratios $f_L$.  Although it is hard to
establish a critical limit where networks can be said to be transitive
or not, the degree of transitivity implied by the connections in a
given complex network can be at least partially quantified in terms of
the completeness ratios $f_L$.  This follows directly from the own
definition of this measurement as the ratio between the number of
edges in the \emph{L-}conditionally expanded network, which correspond to
the number of edges belonging to transitive cycles of length $L+1$,
and the number of edge in the original network.  The completeness
ratios $f_2$ and $f_3$ obtained for the word association experiment
clearly indicate a high degree of transitivity for $L = 2$ and $3$,
with over 50\% of the edges obtained by \emph{2-} and \emph{3-}conditional
expansions belonging to respective cycles.

Because of the high level of connectivity underlying the word
association data, attempts to isolate communities by using \emph{2-} and
\emph{3-}conditional expansions led to a single dominant community and
several rather small clusters.  However, by thresholding the original
weight matrix at $T=2$ instead of $T=1$, i.e. by making
$B=\delta_{T=2}(W)$, a number of interesting communities were obtained
by applying the \emph{3-}conditional expansion over the matrix
$A=\delta_{T=1}(B+B')$.  Table~\ref{tab:comms} shows some of the so
obtained communities, which are characterized by the fact that the
respective constituent words tend to form \emph{4-}cycles such as
\emph{water} $\mapsto$ \emph{drink} $\mapsto$ \emph{soda} $\mapsto$
\emph{cold} $\mapsto$ \emph{water}, therefore leading to longer range
correlations.  The \emph{2-}conditional expansion of $A$ led to just two
communities of three words each.

\begin{table}
  \vspace{1cm}
  \begin{tabular}{||l|r||}  \hline
    1           &  water, drink, cold, soda, ice \\  \hline
    2           &  round, hole, circle, square, table \\   \hline
    3           &  animal, pig, cat, tail \\   \hline
    4           &  Mary, John, man, woman \\   \hline
    5           &  much, good, better, work \\  \hline
  \end{tabular}
\caption{Some of the communities identified by \emph{3-}conditional
expansions applied to the word association experiment.  Observe that
such communities are characterized by the tendency of the respective
words to form \emph{4-}cycles such as \emph{water} $\mapsto$ \emph{drink}
$\mapsto$ \emph{soda} $\mapsto$ \emph{cold} $\mapsto$
\emph{water}.~\label{tab:comms}}
\end{table}

\section{Concluding Remarks}

By introducing and characterizing the concepts of \emph{L-}conditional
expansions, and by showing that successive \emph{L-}percolations can
be associated to a complex network, the current paper has opened a
series of possibilities for further investigations, including not only
the consideration of higher values of $L$ and other evolution models,
but also the use of the introduced concepts as a means to identify
communities in the original network.  As illustrated for the word
association experiment discussed in this work, the introduced
methodology also presents good potential for characterizing real
phenomena.

\begin{acknowledgments}

The author is grateful to FAPESP (proc. 99/12765-2), CNPq
(proc. 308231/03-1) and Human Frontier for financial support.

\end{acknowledgments}

 
\bibliography{high_pre}

\end{document}